\documentclass[submission,copyright,creativecommons]{eptcs}
\providecommand{\event}{International Workshop on Formal Engineering approaches\\ to Software Components and Architectures (FESCA), 2014}

\usepackage{graphics}
\usepackage{subfigure}
\usepackage{tabularx,colortbl}
\usepackage{multirow}
\usepackage{listings}
\usepackage{color}
\usepackage{amsmath}
\usepackage{amssymb}
\usepackage{floatflt}

\lstdefinestyle{acceleo}
  {morekeywords={module,template,public, file, (, ), [/,], if, import, elseif, for, not, and, else},
  sensitive=false,
  morecomment=[l]{//},
  morecomment=[s]{/*}{*/},
  morestring=[b]",
basicstyle=\scriptsize
language=acceleo,                
basicstyle=\footnotesize,       
numbers=left,                   
numberstyle=\scriptsize,        
stepnumber=1,                   
numbersep=5pt,                  
backgroundcolor=\color{white},  
showspaces=false,               
showstringspaces=false,         
showtabs=false,                 
frame=single,	                
tabsize=2,	                    
captionpos=b,
abovecaptionskip=10pt,
belowcaptionskip=0pt,                 
breaklines=true,                
breakatwhitespace=false,        
escapeinside={\%*}{*)},         
}

\lstdefinestyle{drools}
  {morekeywords={when,then,rule, from, accumulate, over, window, entry, point,average,modify,end,dialect, no, loop,salience,declare,@role,@timestamp,this,after, before,coincides,finishes,during,includes,meets},
  sensitive=false,
  morecomment=[l]{//},
  morecomment=[s]{/*}{*/},
  morestring=[b]",
basicstyle=\scriptsize
language=acceleo,                
basicstyle=\footnotesize,       
numbers=left,                   
numberstyle=\scriptsize,        
stepnumber=1,                   
numbersep=5pt,                  
backgroundcolor=\color{white},  
showspaces=false,               
showstringspaces=false,         
showtabs=false,                 
frame=single,	                
tabsize=2,	                    
captionpos=b,
abovecaptionskip=10pt,
belowcaptionskip=0pt,                 
breaklines=true,                
breakatwhitespace=false,        
escapeinside={\%*}{*)},         
}

\title{A model-driven approach to broaden the detection of software performance antipatterns at runtime\thanks{ This work has been partially supported by VISION ERC project (ERC-240555).}}
\author{Antinisca Di Marco, Catia Trubiani
\institute{Dep. of Computer Engineering and Science, and Mathematics\\
University of L'Aquila, L'Aquila, Italy}
\email{\{antinisca.dimarco,catia.trubiani\}@univaq.it}
}
\def\titlerunning{Detection of software performance antipatterns at runtime.}
\def\authorrunning{A. Di Marco, C. Trubiani}

\begin{document}
\maketitle

\begin{abstract}
Performance antipatterns document bad design patterns that have negative influence on system performance. In our previous work we formalized such antipatterns as logical predicates that build on four different views: (i) the static view that captures the software elements (e.g. classes, components) and the static relationships among them; (ii) the dynamic view that represents the interaction (e.g. messages) that occurs between the software elements to provide system functionalities; (iii) the
deployment view that describes the hardware elements (e.g. processing nodes) and the mapping of software resources onto hardware platforms; (iv) the performance view that collects a set of specific performance indices.
In this paper we present a lightweight infrastructure that enables the detection of software performance antipatterns at runtime through the monitoring of specific performance indices. The proposed approach precalculates the logical predicates of antipatterns and identifies the ones whose static, dynamic and deployment sub-predicates occur in the current system configuration and brings at runtime the verification of performance sub-predicates. The proposed infrastructure leverages model-driven techniques to generate probes for monitoring the performance sub-predicates thus to support the detection of antipatterns at runtime.
\end{abstract}

\section{Introduction}\label{sec:intro}

The model-based approaches, pioneered under the name of Software Performance Engineering (SPE)
by Smith \cite{DBLP:conf/cmg/SmithM08}, create performance models and use quantitative results from
these models to adjust the architecture and design \cite{ClementsBachmannEtAl03}
with the purpose of meeting performance requirements \cite{DBLP:conf/icse/WoodsideFP07}.

Several approaches have been successfully
applied by modeling and analyzing the performance of software systems
on the basis of predictive quantitative results
\cite{DBLP:books/daglib/0027475}. However, the problem of interpreting
the performance analysis results (such as mean response time,
throughput variance) is still quite critical, since it is difficult
to translate mean values, variances, and probability distributions
into architectural feedbacks useful to overcome
performance problems (such as split a software component in two components and
re-deploy one of them). Such activities are still exclusively
based on the analysts' experience, and therefore they suffer lack of automation.

Software performance antipatterns \cite{SmithWilliams2003}
have been used to automate the interpretation of performance analysis results and
the generation of architectural feedback \cite{DBLP:journals/jot/ParsonsM08,DBLP:journals/pe/Xu12}.
The rationale of using the performance antipatterns knowledge is two-fold:
on one hand, an antipattern identifies a bad practice in the software architectural model
that negatively affects the performance indices, and on the other hand, its definition also includes
a solution description that lets the software architect devise refactoring actions.
Hence, it is possible to identify from a bad throughput and/or
utilization value the software components and/or interactions responsible for that bad value.

Since the software performance antipatterns
had been originally defined in natural language, in \cite{antipatterns-sosym-2012} we have tackled the problem of providing a less
ambiguous representation.
Performance antipatterns are very complex (as compared to other software patterns) because they
are founded on different characteristics of a software system, spanning from \emph{static} to \emph{behavioral} to \emph{deployment},
and they additionally include values of \emph{performance} indices. This high complexity requires multi-view
representations, as demonstrated in \cite{antipatterns-sosym-2012}.

\begin{floatingfigure}[l]{9.5cm}
	\centering
\resizebox{0.6\textwidth}{!}{%
  \includegraphics{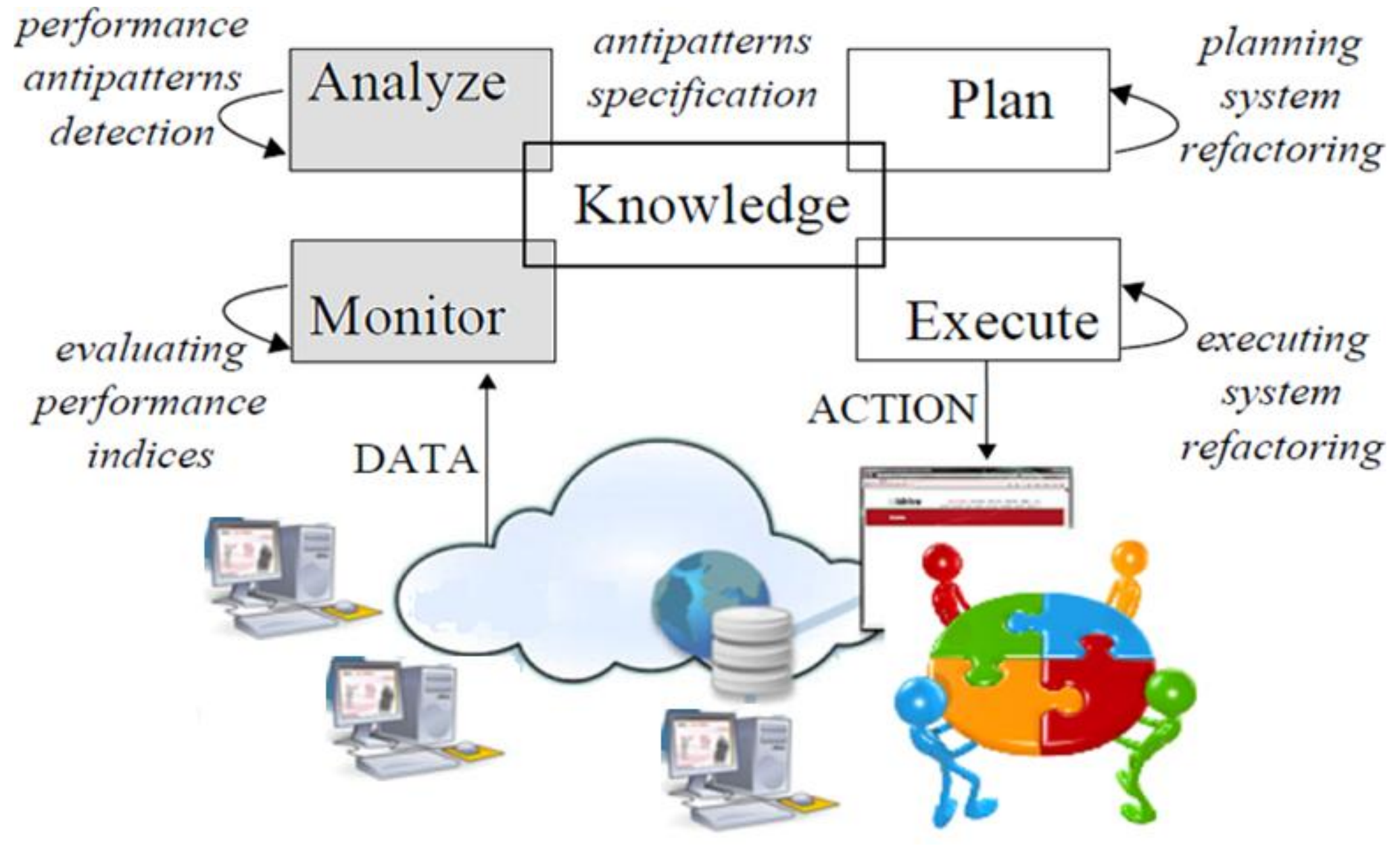}}
	\caption{Antipattern-based MAPE process.}
	\label{fig:antipat-MAPE}
\end{floatingfigure}

In this paper we move a further step ahead, in that we undertake the problem of detecting performance antipatterns
at runtime.
Inspired by the IBM autonomic control loop \cite{mape-ibm}, Figure \ref{fig:antipat-MAPE}
schematically represents the operational steps of the antipattern-based reconfiguration process we propose.
The overall process of \textbf{M}onitoring, \textbf{A}nalyzing, \textbf{P}lanning,
and \textbf{E}xecuting (MAPE) is executed to assess and, if needed, improve the performance
properties of a software architecture under development.
Software architectures are characterized by a huge amount of interactions among software
elements; such interactions are input to the antipattern-based process whose goal is to
provide and make use of \emph{knowledge}.
Shaded boxes of Figure \ref{fig:antipat-MAPE} represent the focus of this paper.
In particular, our approach supports two main activities: (i) \emph{Monitor},
i.e. the monitoring of performance indices; (ii) \emph{Analyze}, i.e. the detection
of performance antipatterns. After detecting antipatterns, there are two
activities that must be executed to complete the feedback process: (i) \emph{Plan},
i.e. the antipattern-based reconfiguration actions are planned; (iv) \emph{Execute},
i.e. the system is reconfigured by removing the detected antipatterns.

The goal of this paper is to broaden the detection of software performance antipatterns at runtime
by introducing a model-driven monitoring infrastructure able to measure a specific set of performance
indices. To this end, we introduce a model-based approach that allows
to precalculate the logical predicates of performance antipatterns and
identifies the ones whose static, dynamic and deployment sub-predicates occur
in the system configuration under analysis. The proposed infrastructure leverages model-driven
techniques to generate probes for the verification of performance sub-pre\-di\-ca\-tes
by monitoring a sub-set of performance indices only.

The paper is organized as follows.
Section \ref{sec:PA} provides some background information on software
performance antipatterns. The process for the detection of antipattern at runtime is described in Section \ref{sec:process}.
Section \ref{sec:infrastructure} presents the monitoring infrastructure we use to
verify performance sub-predicates.
Section \ref{sec:casestudy} shows the approach at work on a case study.
Section \ref{sec:related} discusses the related works,
and finally in Section \ref{sec:conclusion} conclusions and future
research directions are sketched.

\section{Performance Antipatterns}
\label{sec:PA}

\begin{figure}[htbp]
	\centering
\resizebox{1\textwidth}{!}{%
  \includegraphics{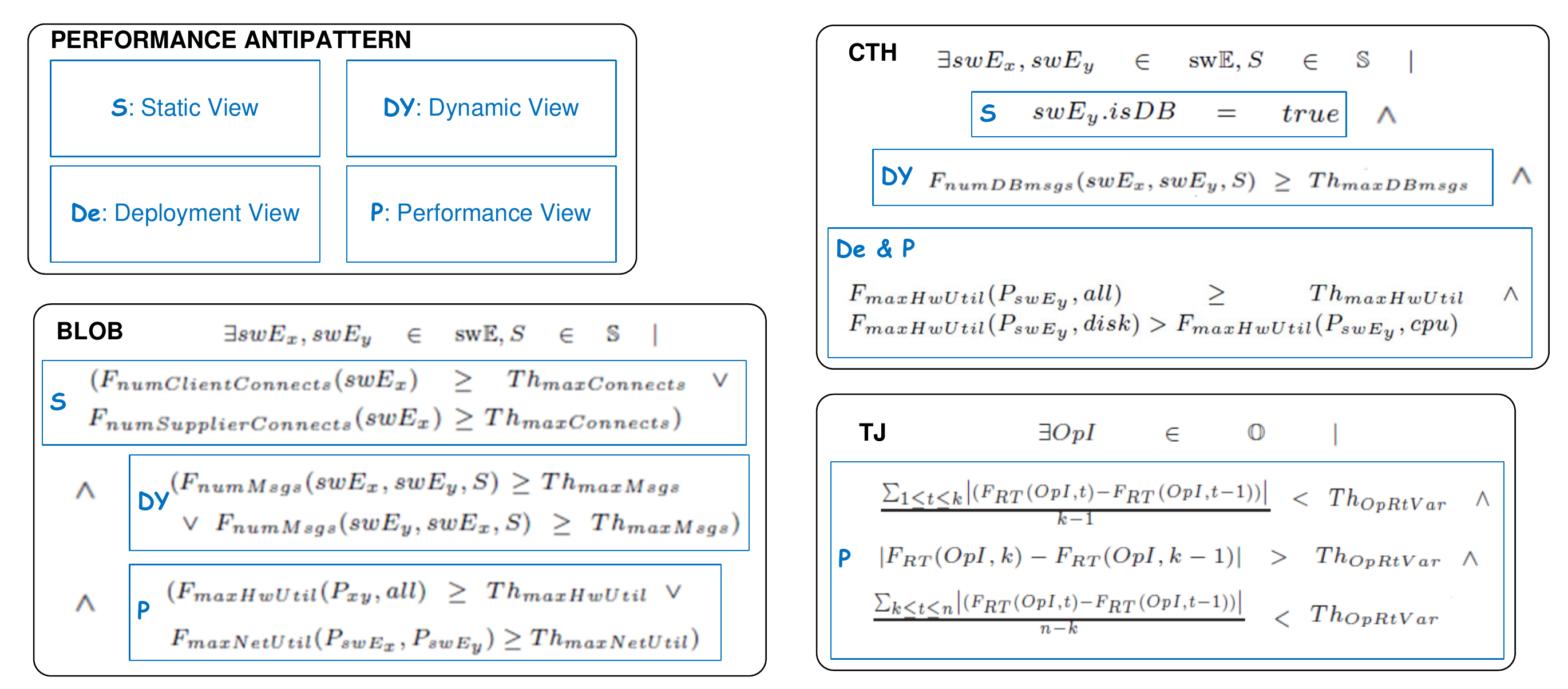}}
	\caption{Performance antipatterns modeling.}
	\label{fig:PAmodeling}
\end{figure}

Figure \ref{fig:PAmodeling} reports the formalization we provided in \cite{antipatterns-sosym-2012}
for  Blob, the Circuitous Treasure Hunt (CTH), and the Traffic Jam (TJ) antipatterns.
As mentioned in Section \ref{sec:intro}, antipatterns are founded on different characteristics of
a software system that we organized in four different views: (i) the \emph{static view} \textbf{S} that
captures the software elements (e.g. classes, components) and the static relationships among them;
(ii) the \emph{dynamic view} \textbf{Dy} that represents the interaction (e.g. messages) that occurs between
the software elements to provide the system functionalities; (iii) the \emph{deployment view} \textbf{De} that
describes the hardware elements (e.g. processing nodes) and the mapping of software resources onto
hardware platforms; (iv) the \emph{performance view} \textbf{P} that collects a set of specific performance indices.

\vspace{-0.20cm}
\vspace{-0.10cm}

To determine when performance antipatterns occur, our specification includes thresholds on design (i.e. static, dynamic and deployment)
features (e.g. \emph{many} usage dependencies, \emph{excessive} message traffic) and performance indices
(e.g. \emph{high}, \emph{low} utilization).

For example, a Blob occurs when a component requires a \emph{lot} of information from other ones,
it generates \emph{excessive} message traffic that lead to \emph{over utilize} the device on which
it is deployed or the network involved in the communication. Its logic-based representation
is reported in Figure \ref{fig:PAmodeling} where four thresholds have been
defined:
(i) $Th_{maxConnects}$ indicates the maximum number of connections a component may be involved in;
(ii) $Th_{maxMsgs}$ indicates the maximum number of messages a component may send in an interaction;
(iii) $Th_{maxHwUtil}$ indicates the upper bound for processing device utilization;
(iv) $Th_{maxNetUtil}$ indicates the upper bound for network device utilization.
The first two thresholds refer to static and dynamic features, respectively, whereas the latter ones refer to
performance indices.
As another example, a TJ occurs when one problem causes a backlog of jobs that results in a wide
variability in response time. Its logic-based representation is reported in Figure \ref{fig:PAmodeling} where one thresholds has
been defined: (i) $Th_{OpRtVar}$ indicates the maximum bound for the variability
in response times of operations across time intervals, and it refers to a performance index.

Each antipattern is formalized by a set of sub-predicates referring to the defined four views, as depicted in Figure \ref{fig:PAmodeling}. Note that there are some antipatterns whose predicates span on all the four views (such as CTH), some others referring to a sub-set of views (such as Blob) and finally there are some antipatterns whose predicates refer to the performance view only (such as TJ). Besides the TJ,
there are three other antipatterns (over a set of twelve antipatterns we formalized in \cite{antipatterns-sosym-2012})
that refer to the performance view only in their modeling, i.e. 'Concurrent Processing System' (CPS), 'The Ramp', and 'More is Less'.
For more details on these antipatterns please refer to \cite{antipatterns-sosym-2012}.

\section{Performance Antipattern Detection: Process at Runtime} \label{sec:process}

In this section we describe the process we envisage for the detection of performance
antipatterns at runtime.
To speed up the detection activity, our process is constituted by two main
operational steps: (1) a full automatic \emph{pre-calculus} of antipatterns that off-line determines the set of antipatterns instances for which the current running configuration satisfies the static, dynamic, and deployment views; (2) \emph{monitoring} of performance indices for the set of antipatterns instances identified during the pre-calculus step. In this way the effort spent by the monitoring is reduced since we perform at runtime the verification of a, possibly limited, sub-set of antipattern instances only.

We recall that there are four antipatterns whose logical formula includes the performance view predicates only
(see Section \ref{sec:PA}), hence the pre-calculus step is not executed for them. The detection of these antipatterns
is due to a set of performance indices that must be always monitored at runtime.

\begin{figure}[htbp]
	\centering
\resizebox{0.7\textwidth}{!}{%
  \includegraphics{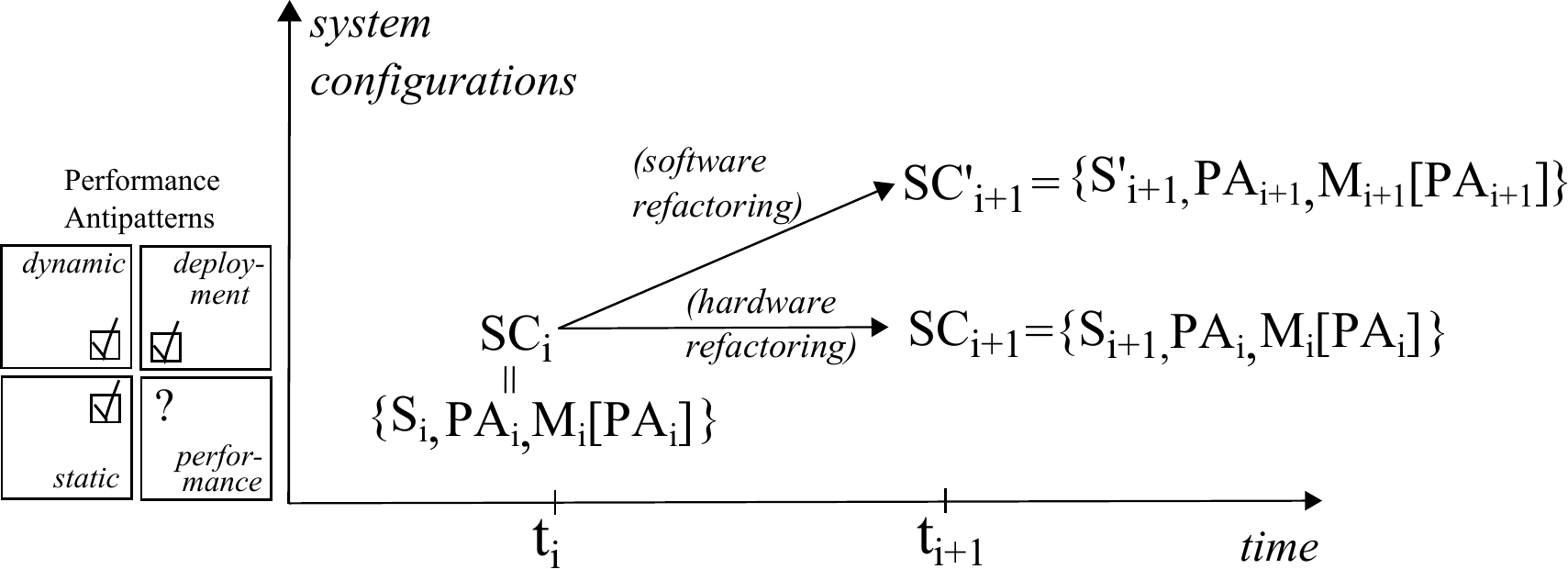}}
	\caption{System Configuration, Antipatterns Instances and Active Monitors over time.}
	\label{fig:SC-overTime}
\end{figure}

Figure \ref{fig:SC-overTime} illustrates how the system configuration ($SC$) is modified over time while
detecting and solving performance antipatterns. In a generic instant of time $t_i$ a system configuration $SC_i$ is
constituted by three elements: \{$S_i$, $PA_i$, $M_i$[$PA_i$]\}, where $S_i$
represents the running system, $PA_i$ is the set of antipattern instances whose
(static, dynamic, deployment) views have been already verified in $S_i$, and $M_i$
represents the active monitors generated on the basis of $PA_i$ that must check the verification of their performance view predicates.

The detected antipatterns constitute the knowledge on which refactoring actions are devised. In particular,
two different types of refactorings can be considered at a generic instant of time $t_i$: software and hardware refactorings.
\emph{Software refactoring} generates a new system configuration ($SC'_{i+1}$) where
the running configuration is modified ($S'_{i+1}$) and the pre-calculus step returns a new set of instances ($PA_{i+1}$) together with a new set of active monitors $M_{i+1}$[$PA_{i+1}$].
Some examples of software refactoring are: (i) the splitting of software components, (ii) the redeployment of
software components into different hardware platforms, (iii) the replacement of software components with ones
that require lower amounts of resource demands, etc.
By \emph{hardware refactoring} we consider all the refactoring actions that do not modify the (static, dynamic, deployment)
characteristics of software systems thus to skip the pre-calculus step since the set of antipatterns and active monitors
do not change. Some examples of hardware refactoring are: (i) the increasing of the processing power of processors, disks,
and/or network resources, (ii) the increasing of the multiplicity of processors, disks, and/or network resources, etc.
Note that software/hardware refactorings may incur different costs thus inducing a quality trade-off issue\footnote{For example the replacement of software components
is enabled by different implementations of the same components that may require other amounts of resource demands
but it may have different development and/or monetary costs.}. For the sake of simplicity,
we do not consider this issue here.

\section{Property-Driven Monitoring Infrastructure} \label{sec:infrastructure}

In this section we describe the monitoring infrastructure that evaluates the antipatterns performance view.
We recall that this latter view is represented by a set of performance properties the running system must manifest
during its execution to detect the corresponding antipattern. Such predicates are generally defined for each antipattern
(see Section \ref{sec:PA}) but must be actualized for the specific system under analysis. For example, the TJ antipattern
 occurs under specific conditions related to the response time ($F_{RT}$) of one operation instance ($OpI$) at a given time $t$.
 The actual detection of the TJ antipattern at runtime includes the specification of the \emph{generic} $OpI$ that, customized
 for a specific system, become the \emph{actual} operation instance that needs to be monitored.

The monitoring infrastructure we propose takes advantage of model-driven techniques since the set of performance antipatterns
to detect (and the corresponding properties to monitor) can change at runtime (see Section \ref{sec:process}). In particular,
the process must be iterated each time the properties to be monitored change, hence a substantial human effort and specialized
expertise is required if the high level description of system properties must be translated into lower-level monitor directives.
Our model-based infrastructure builds upon a property-driven monitoring framework we proposed in \cite{SERENE11},
and it automatically translates the predicates of antipatterns performance view into a concrete monitoring setup.

The sequel of the section discusses the performance view properties modeling approach in
Section \ref{sec:modeling}, and presents the configurable monitoring infrastructure we propose for the monitoring activity
in Section \ref{sec:monitoring}.

\subsection{Performance view properties modeling approach}\label{sec:modeling}

\begin{figure}[htbp]
\vspace{-0.2cm}
	\centering
\resizebox{\textwidth}{!}{%
  \includegraphics{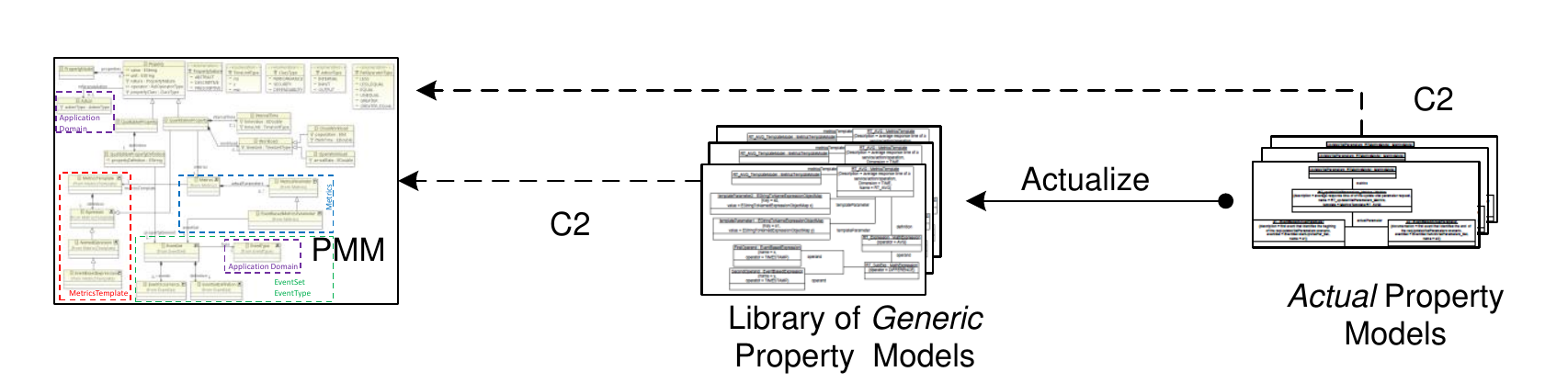}}
	\caption{Performance View Properties Modeling Approach.}
	\label{fig:modeling}
\end{figure}

Figure \ref{fig:modeling} illustrates our performance view properties modeling approach.
Both Generic and Actual properties are modeled by using the Property Meta-Model (PMM) \cite{QASBA11,QUATIC12},
and the corresponding models are conform to it (\emph{C2} relation in Figure \ref{fig:modeling}). Generic Property
Models are defined once and are generic w.r.t. the events to observe, whereas Actual Property Models, being
heavily-coupled with the specific system to monitor, are specified time by time when the thresholds and the events
to be observed are determined and the running system configuration ($SC_i$) is known.

\begin{figure}[!t]
\vspace{-0.8cm}
	\centering
\resizebox{1\textwidth}{!}{%
  \includegraphics{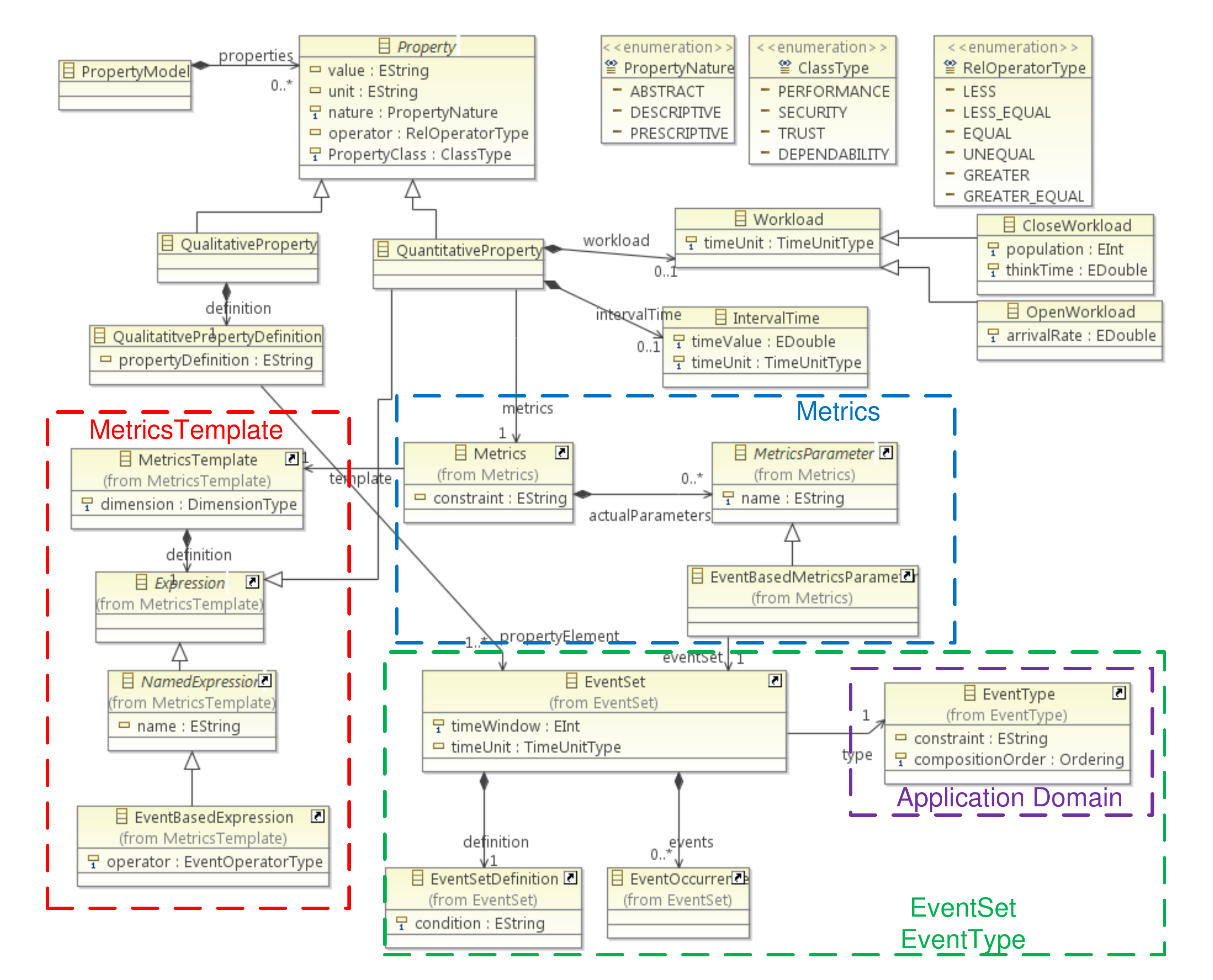}}
  \vspace{-0.8cm}
	\caption{PMM structure.}
	\label{fig:PMM}
\vspace{-0.1cm}
\end{figure}

Note that the Generic Property Models represent a pre-defined Library we built from the logic-based representation of antipatterns discussed in Section \ref{sec:PA},
and the Actual Properties Models are obtained by actualizing the Generic Property Models (\emph{Actualize} relation in Figure \ref{fig:modeling}) to the specific system.

The actualization step is not fully automatic: while the events to observe are provided by the pre-calculus, the thresholds on performance indices must be initially set by experts and, possibly, dynamically adapted by considering specific heuristics on observed performance values.

\paragraph*{Property Meta-Model}
Figure \ref{fig:PMM} reports the PMM key concepts and their relationship.
PMM includes \emph{PRESCRIPTIVE} and \emph{DESCRIPTIVE} properties of systems.
A property can be qualitative (\emph{QualitativeProperty}) or quantitative (\emph{QuantitativeProperty}). In general, quantitative properties are related to performance or dependability, whereas \emph{QualitativeProperty} refers to properties about the event occurrences that are observed and cannot be measured. They in general refer to the behavioral description of the system (e.g. deadlock freeness or liveness). Quantitative properties are measurable and have associated \emph{Metrics}.
A \emph{QuantitativeProperty} can be associated to a \emph{Workload} and/or an \emph{IntervalTime}.

\emph{MetricsTemplate} models a generic metric (e.g. the response time as the duration of an event/operation), the concrete \emph{Metrics} that refers to a \emph{MetricsTemplate} and specifies the (actual) \emph{metricParameter} for the (formal) \emph{templateParameter} in the application domain the metric has been defined. The specification of the concrete event in the metric is done via the \emph{EventType} the metric refers to. Such \emph{EventType} refers to observable operations or events belonging to the Application Domain for which the \emph{Property} needs to be verified/guaranteed. An \emph{EventSet} represents a set of event instances the refer to the same \emph{EventType}.

PMM is implemented as an eCore model using Eclipse Modeling Framework (EMF) \cite{EMFurl} and it is provided with an associated editor realized as an Eclipse Plugin. This editor allows to create new model instances of the \emph{Property}, \emph{Metrics}, \emph{MetricsTemplate}, \emph{EventType} and \emph{EventSet} meta-models.

\vspace{-0.2cm}

\begin{figure}[htbp]
	\centering
\resizebox{1\textwidth}{!}{%
  \includegraphics{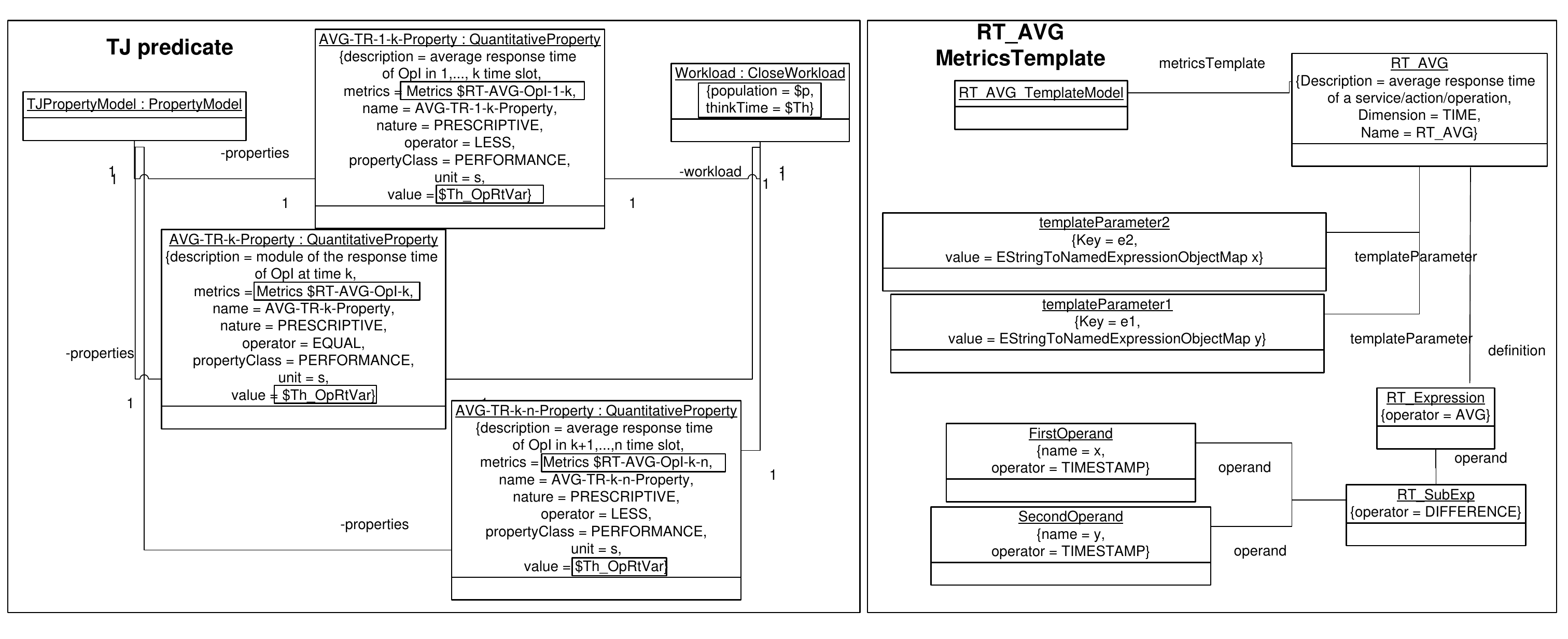}}
	\vspace{-0.8cm}
	\caption{Traffic Jam Performance Predicate and Response Time Metrics Template.}
	\label{fig:rt_template}
\end{figure}

\paragraph*{Library of Generic Property Models}
Figure ~\ref{fig:rt_template} shows an example of a generic property model in the library, namely TJPropertyModel, modeling the \emph{TJ} antipattern performance predicate (at the left-hand of the figure) and the average response time \emph{MetricsTemplate} used in it (at the right-side of the figure).

The \emph{TJPropertyModel} contains three \emph{QuantitativeProperty} models, one for each term of the performance predicate. The three QuantitativeProperty models refer to the same property on different time slots, hence for the sake of space, we only describe the \emph{AVG-TR-1-k-Property}. This is a PERFORMANCE PRESCRIPTIVE property with two parameters, \$RT-AVG-OpI-1-k and \$Th\_OpRtVar. The former refers to the metrics to be used for the specific operation, whereas the latter parameter indicates the threshold for the average response time of the operation.  Furthermore, AVG-TR-1-k-Property has a generic closedWorkload with two parameters to  be actualized: \$p representing the population of the closedWorkload and \$Th the think time of each individual of the population. To obtain an Actual Property Model for the TJ antipattern, all the parameters of the \emph{TJPropertyModel} need to be actualized to a concrete scenario.

$RT\_AVG$ MetricsTemplate represents a \emph{TIME} measure defined as average of the
differences for the timestamps of two generic event instances ($x$ and $y$ in the model).
The template exposes two \emph{templateParameters}: $e_1$ bound to $y$, and $e_2$ bound to $x$.
A Metrics, whose definition is an instance of the MetricsTemplate,
concretises the template for a specific scenario.
This is reflected in the metrics' actual parameters
which substitute the template parameters by linking the general
description to the specific application domain the metrics refers to.
We will see in Section \ref{sec:casestudy} the \emph{RT\_updateVitalParameters\_Metrics} that
actualizes the corresponding $RT\_AVG$ MetricsTemplate by linking to the templateParameters $e_1$ and $e_2$,
two EventSets ($startUpVitalPar\_Set$ and $fwAckVitalParameters\_Set$ respectively), and specifying that
the two event sets must satisfy a metrics constraint (i.e. the two event sets must be related to each other).

\subsection{Configurable Monitoring Infrastructure for Antipattern Detection}\label{sec:monitoring}

\begin{figure}[htbp]
\vspace{-0.8cm}
	\centering
\resizebox{0.9\textwidth}{!}{%
  \includegraphics{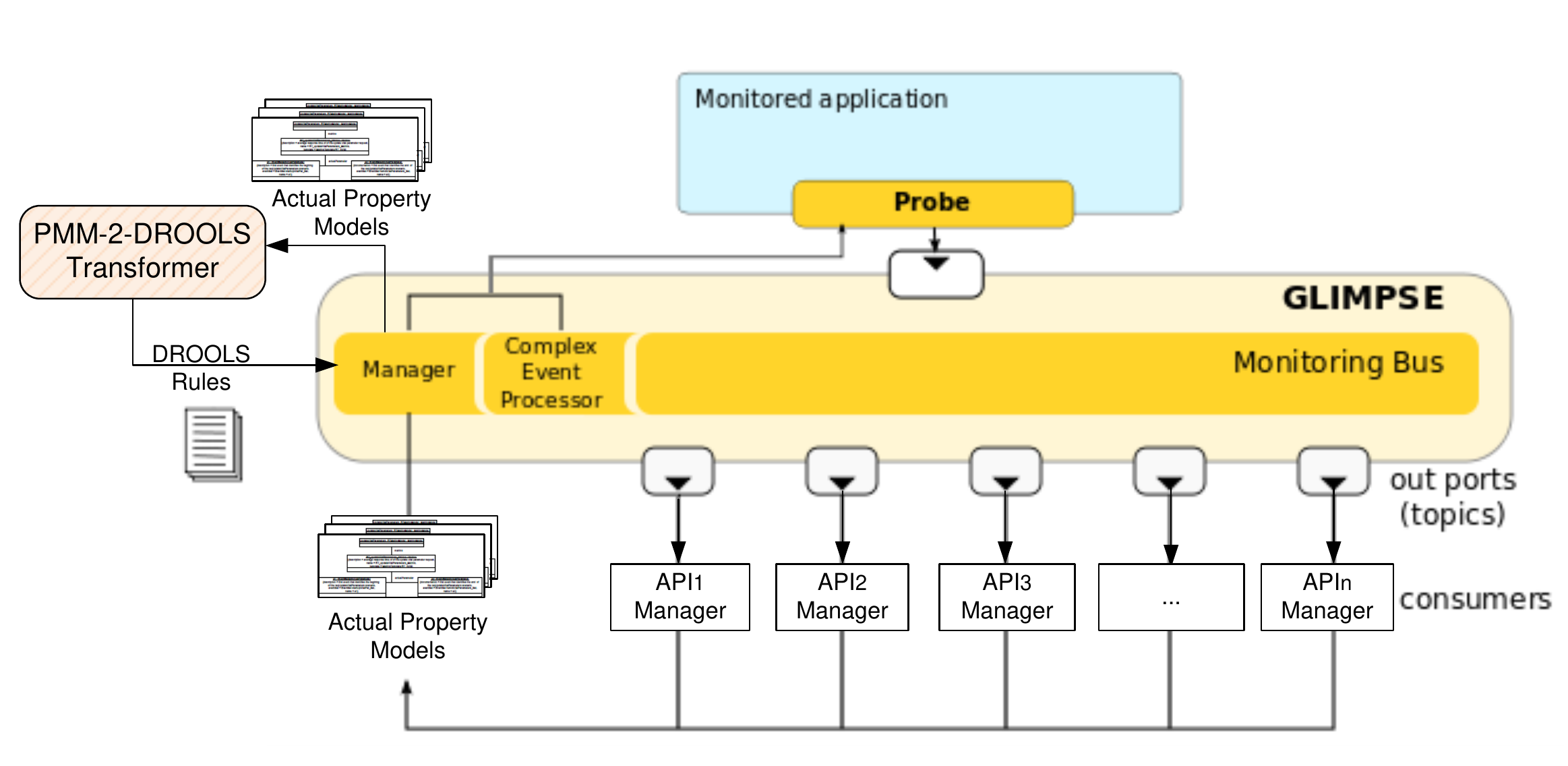}}
\vspace{-0.4cm}
	\caption{Property-Driven Monitoring Infrastructure for Antipattern Detection.}
	\label{fig:configuration}
\vspace{-0.1cm}
\end{figure}

Figure \ref{fig:configuration} depicts our configurable monitoring infrastructure. It is composed by: \emph{i)} a generic monitoring framework (i.e. GLIMPSE \cite{ewdc11}), and \emph{ii)}  a  PMM-2-Drools Translator that generates DROOLS rules from PMM Actual Property Models, and such rules are used to configure GLIMPSE to the actual properties to monitor at run time.

As shown Figure \ref{fig:configuration}, the GLIMPSE manager component takes as input the property model and activates an external component that performs the code generation according to Drools Fusion complex event processor \cite{droolsFusion}. The output of this transformation is represented by a specific rule that is processed by the complex event processor component of GLIMPSE.
In this way, we configure the generic framework GLIMPSE to a concrete monitoring infrastructure.

\paragraph*{Generic fLexIble Monitoring based on a Publish-Subscribe infrastructurE (GLIMPSE)}
As shown in Figure \ref{fig:configuration} GLIMPSE, developed with the goal of decoupling the event specification from the analysis mechanism, is composed of five main components:

\noindent \emph{Probes} that intercept primitive events when they occur in the software and send them to the GLIMPSE Monitoring Bus. Probes are usually realized by injecting code into an existing software or by using proxies. The generated Drools rules are used from the Manager to instruct the probes for monitoring the system characteristics of predicates belonging to the antipatterns performance view. Note that the probes instrumentation strictly depends on the Actual Property ($AP_i$) precalculated for the system configuration $SC_i$ (see Section \ref{sec:process}).

\noindent \emph{Monitoring Bus} that is the communication backbone that all information (events, questions, answers) is sent on: Probes, Consumers, Complex Event Processor and by all the services querying information to GLIMPSE. A publish-subscribe paradigm devoting the communication handling to the Manager component is here adopted.

\noindent \emph{Complex Event Processor}  (CEP) is the rule engine which analyzes the primitive events, generated from the
probes, to infer complex events matching the consumer requests. There are several rule engines that can be used for this task. The current GLIMPSE implementation adopts the Drools Fusion rule language~\cite{droolsFusion}.

\noindent \emph{Consumer} that is a simple customer requiring some information to be monitored. It sends a request (Actual Property Model) to the Manager using the Monitoring Bus and waits for the evaluation results on a dedicated channel provided by the Manager. Consumers are software components that manage the antipattern instances to detect and solve. In particular, we have one AP Manager for each antipattern instance in $PA_i$  of the system configuration $SC_i$ (as defined in Section \ref{sec:process}).

\noindent \emph{Manager} that is the orchestrator of the GLIMPSE architecture. It manages the communications among the GLIMPSE components.
Specifically, the Manager fetches requests coming from Consumers, analyzes them and instructs the Probes.
Then, it instructs the CEP Evaluator, creates and notifies to the Consumer a dedicated channel on which it will provide results produced by the CEP Evaluator.

\paragraph*{PMM Model to DROOLS Rules Translator}
The translation of the PMM model into Drools rules automatizes the GLIMPSE configuration transforming the PMM models into a concrete monitoring setup. The Model2Code Transformer component has been implemented using the Acceleo code generator IDE \cite{acceleo} that supports the automated code generation from UML and EMF and uses the \emph{template}s to generate code (or text) from a model.

The implemented Model2Code Transformer has a \emph{.mtl} extension, it takes as inputs the PMM ecore models (specifically Core.ecore, EventSet.ecore, EventType.ecore, Metrics.ecore, MetricsTemplate.ecore, Property.ecore) and the model (compliant to PMM) of the property, metrics or event to be monitored, and produces as output one or more Drools rules, that are provided in input to the GLIMPSE complex event processor.

The Transformer consists of a main generation rule or template that calls a certain number of other templates, each one is applied on a PMM model element (or object) to produce some text. In each template there are static areas, and they will be included in the generated file as they are defined, whereas dynamic areas correspond to the expression evaluation on the current object.


\vspace{0.2cm}
\begin{lstlisting}[style=acceleo,
                   	caption={Main Template of Model2Code Transformer},
		   			label={lst:transformerMain}, breaklines,
		   			frame=single,
		   			xleftmargin=15pt]
[comment encoding = UTF-8 /]
[module generate('cpmm/model/Core.ecore','cpmm/model/EventType.ecore','cpmm/model/EventSet.ecore','cpmm/model/Metrics.ecore','cpmm/model/MetricsTemplate.ecore','cpmm/model/Property.ecore')]
[import cpmm::acceleo::utilities::utility]
[template public generateElement(p : Property)]
[comment @main/]
[file (p.name, false, 'Cp1252')]
 [if p.oclIsTypeOf(QuantitativeProperty)]
  [processMainQuantitativeProperty(p)/]
 [elseif (p.oclIsTypeOf(QualitativeProperty))]
  [processQualitativeProperty(p)/]
 [/if]
[/file]
[/template]
\end{lstlisting}

As showed in Listing \ref{lst:transformerMain}, the heading section defines which meta-model the template will apply for and the generated file, and a main template (called \emph{generateElement}) is defined to represent the main generation rule. It takes as input the property model and if this model is about a quantitative property then a specific template named \emph{processMainQuantitativeProperty} is  called, otherwise the \emph{processQualitativeProperty} is called.
The \emph{processMainQuantitativeProperty} calls a specific template processing the associated metrics. In this template the
Metrics model is navigated for identifying the associated MetricsTemplate model, then the MetricsTemplate parameters, the associated EventBasedExpressions and their operators. These EventBasedExpressions represent expressions based on events and their name represents the observable, simple or complex, event/behavior the operator applies to.

\section{Case Study: E-Health System}
\label{sec:casestudy}

The E-Health System (EHS) supports the doctors' everyday activities, such as the
retrieval of information of their patients.
On the basis of such data they may send an alarm in case of warning conditions.
The patients are allowed to retrieve information about the doctor expertise and they
update some vital parameters (e.g. heart rate) thus to provide the knowledge
aimed at monitoring their health.
The service we analyze in the following is the \emph{UpdateVitalParameters},
since it is required by a large number of users that originate a huge number
of processes at approximately the same time.

\begin{figure}[htbp]
    \centering
        \resizebox{1.0\textwidth}{!}{%
        \includegraphics{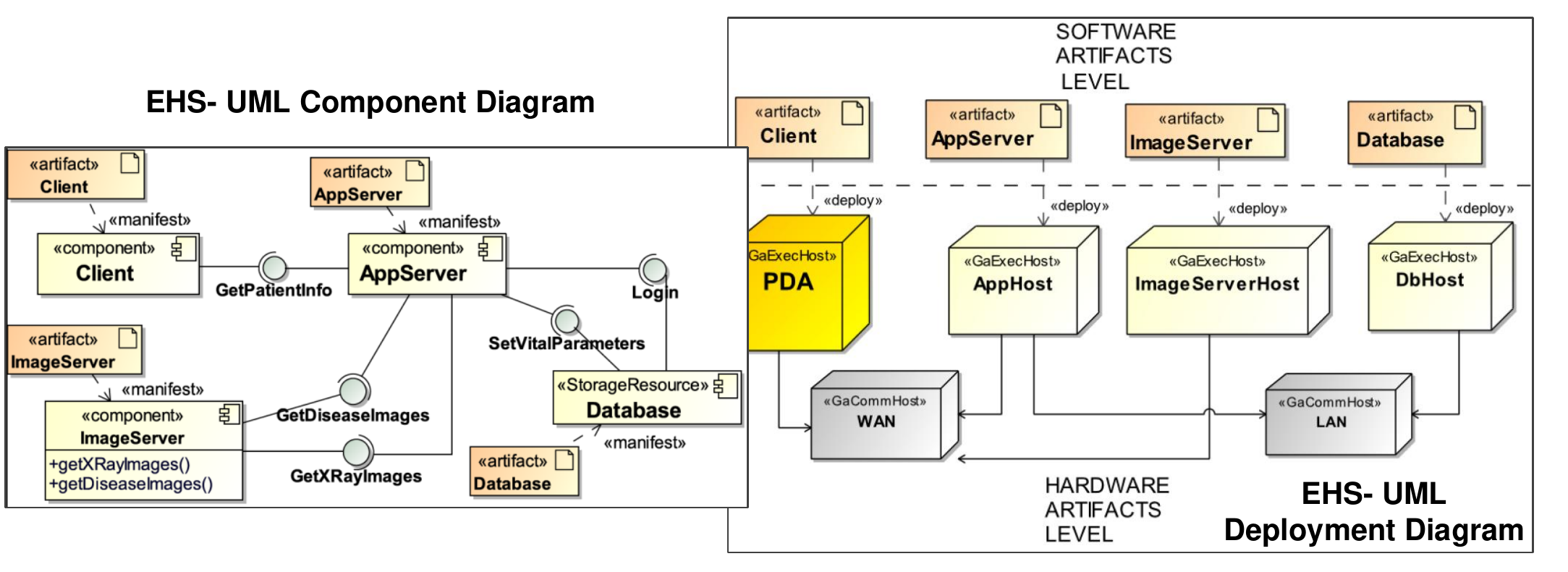}}
        \vspace{-0.8cm}
    \caption{EHS- UML Component and Deployment Diagram.}
  \label{fig:caseStudy:ehs-componentDiagram}
\end{figure}

To conduct this proof of concept, we adopt a model-based performance analysis.
However, the presented approach remains feasible for any running system.
We model EHS with UML \cite{UML2} profiled with MARTE \cite{MARTE}.
Figure \ref{fig:caseStudy:ehs-componentDiagram} reports an excerpt of the EHS component and deployment diagrams
illustrating the software components and their interconnections, and the deployment of the software components
on the hardware platforms. The client application has been deployed on a
Personal Digital Assistant (PDA), i.e. a mobile device in the hands of patients.

\begin{figure}[htbp]
    \centering
        \resizebox{0.85\textwidth}{!}{%
        \includegraphics{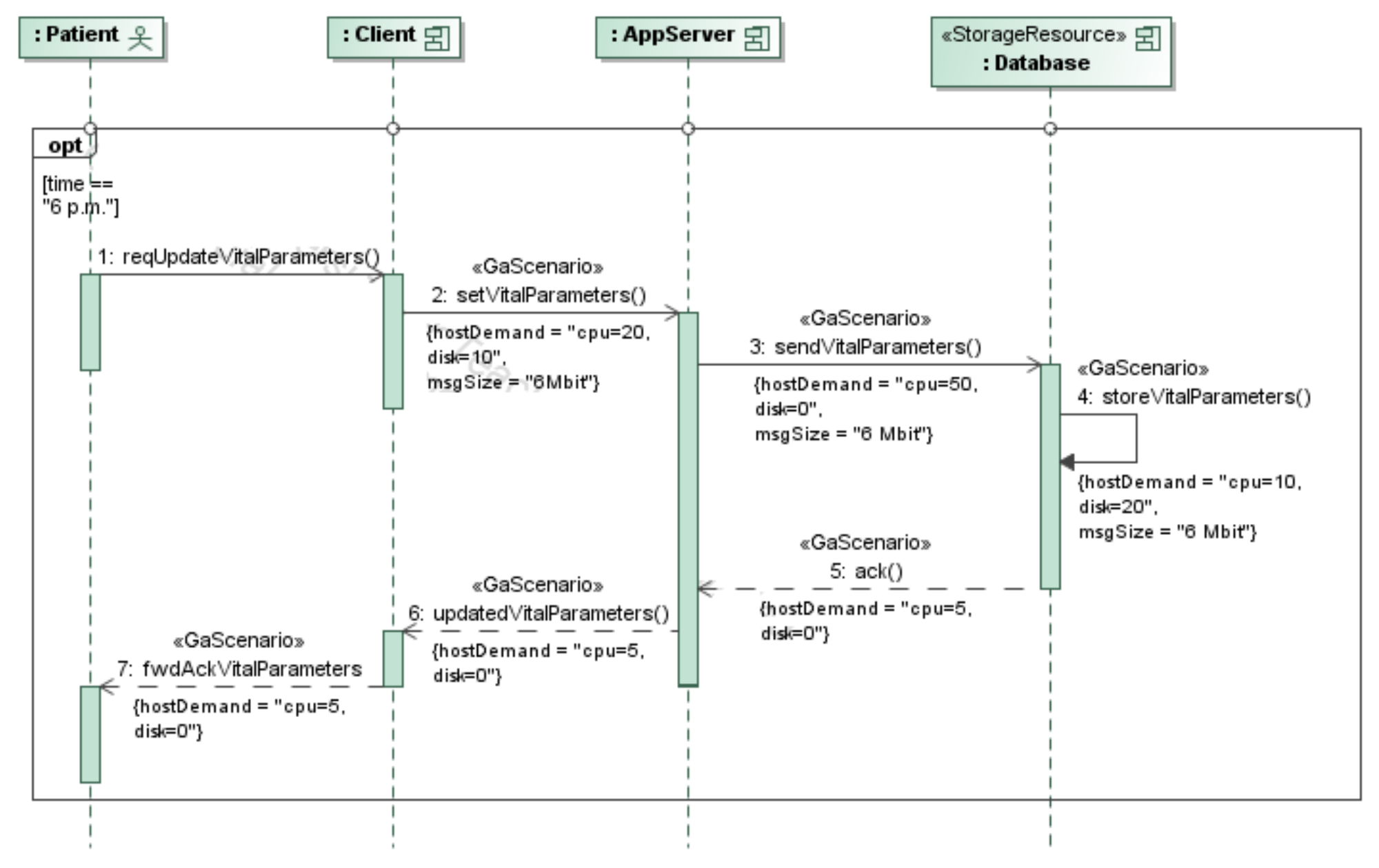}}
        \vspace{-0.2cm}
    \caption{EHS- UML Sequence Diagram for the \emph{UpdateVitalParameters} service.}
  \label{fig:caseStudy:ehs-sequence-updateVitalParameters}
\end{figure}

Figure \ref{fig:caseStudy:ehs-sequence-updateVitalParameters} reports the sequence diagram
for the \emph{UpdateVitalParameters} service.
It describes the communication between software components for the \emph{UpdateVitalParameters} service:
a \emph{opt} fragment indicates that at 6 p.m. patients need to update their vital parameters hence
their client application sends a request to the application server that is meant to
manage the communication towards the database. Vital parameters of patients are sent to the database,
stored on it and, afterwards such update is notified to the client and visualized to the patient.

\vspace{-0.40cm}

The sequence diagram of Figure \ref{fig:caseStudy:ehs-sequence-updateVitalParameters}
contains annotations that will be used in the model-based performance analysis.
For example, the \emph{setVitalParameters} is annotated with the MARTE stereotype
\emph{GaScenario} through which it is possible to specify the usage of hardware resources:
in particular it requires: (i) 20 instructions from a cpu device, and
(ii) 10 accesses from a disk device. The workload addressed to hardware
resources is calculated as the sum of demands, hence in this case the patient's device
executes $25$ cpu instructions (20 for \emph{setVitalParameters} plus $5$ for
\emph{fwdAckVitalParameters}) and $10$ disk accessed (10 for \emph{setVitalParameters}
plus 0 for \emph{fwdAckVitalParameters}). Consider as another example of annotations
that the communication between the patient's device and the \emph{AppHost} requires to exchange
over the network a message of size equal to $6$ Mbit.

A performance requirement has been defined on the RT(\emph{Up\-da\-te\-Vi\-tal\-Pa\-ra\-me\-ters})
service, i.e. such service has to be offered in less than 0.5 seconds, while considering a workload
of ten thousands patients with a thinking time equal to 86400 seconds, i.e. patients update their vital
parameters each day.

The model-based performance analysis phase has been conducted as follows.
We used the Prima-UML methodology \cite{primaUML} to automatically
derive a Queueing Network (QN) \cite{qnBook-Kleinrock}
starting from the UML models related to the \emph{UpdateVitalParameters} service.
The QN has been analyzed with JMT \cite{DBLP:conf/wosp/CasaleS11}.

\vspace{-0.1cm}

The obtained QN model has a set of queueing centers (e.g. \emph{cpu-PDA})
representing the hardware resources of the system, and a set of delay centers
(e.g. \emph{WAN}) representing the network delays.
The input parameters adopted for this experiment are reported in Table \ref{tb:inputParamsEHS}:
the first column lists the QN service centers; the second column
shows the average service demands for the class of job under analysis.

We have referred to \cite{book-ComputerArchitecture} for typical performance
parameters, like processors, disks and networks latencies. For example,
in Table \ref{tb:inputParamsEHS} we can see that the \emph{cpu-PDA} queueing service center requires 0.00375 ms
and it is obtained from $0.000015*25$ in which 0.000015 ms represents the service time of a microprocessor multiplied for the number of
cpu instructions (25, as stated in the sequence diagram of Figure \ref{fig:caseStudy:ehs-sequence-updateVitalParameters}).
Network communication links are subjected to the message size, in fact in Table \ref{tb:inputParamsEHS}
we can see that the \emph{WAN} delay center requires 500 ms
and it is obtained from $6/0.012$ in which 6 Mbit represents the size of the message
exchanged over the network (see Figure \ref{fig:caseStudy:ehs-sequence-updateVitalParameters})
and it is divided for the service time of a wired area network (12 Mbit/sec).

\begin{table}[ht]
\scriptsize
\begin{minipage}[b]{0.5\linewidth}
\centering
{\scriptsize
 \begin{tabular}{|l||c|}
            \hline
             Service Center & Input parameters \\
            \hline
            \hline
             cpu-PDA & 0.00375 ms\\
             \hline
             disk-PDA & 57 ms \\
             \hline
             WAN & 500 ms \\
             \hline
             cpu-AppHost & 0.00825 ms \\
             \hline
             LAN & 60 ms \\
             \hline
             cpu-DbHost & 0.00225 ms \\
             \hline
             disk-DbHost & 114 ms \\
             \hline
\end{tabular}
}
\caption{EHS- QN model input parameters.}
\label{tb:inputParamsEHS}
\end{minipage}
\begin{minipage}[b]{0.5\linewidth}
\centering
{\scriptsize
  \begin{tabular}{| l | l | c |}
    \hline
    antipattern & parameter & value \\ \cline{2-3}
    \hline
    \hline
    & \emph{\$Th\_maxConnects} & 4 \\
    Blob & \emph{\$Th\_maxMsgs} & 5 \\
    & \emph{\$Th\_maxHwUtil} & 0.8 \\
    & \emph{\$Th\_maxNetUtil} & 0.7 \\ \cline{1-3}
    & \emph{\$Th\_initSlot} & 0 \\
    TJ & \emph{\$Th\_sizeSlot} & 50 \\
    & \emph{\$Th\_endSlot} & 1500 \\
    & \emph{\$Th\_OpRtVar} & 0.3 \\
    \hline
  \end{tabular}
 }
\caption{EHS- parameters binding.}
\label{tb:perfParamsBinding}
\end{minipage}
\end{table}

Our model-based performance analysis predicts that the system offers the
\emph{UpdateVitalParameters} service with an average time of 0.61 seconds
that is greater than the required 0.5 seconds. Hence, the unfulfillment
of the requirement triggers the detection of performance antipatterns.

Antipattern-based detection rules contain parameters that
need to be actualized on the analyzed model and basically represent thresholds
that formalize system features. We provide heuristics
to estimate their numerical values. For example, Table \ref{tb:perfParamsBinding}
report the thresholds we consider for Blob and TJ antipatterns. Such values allow to
actualize the parameterized detection rules into instantiated
ones, thus proceeding with the actual detection.
The TJ antipattern requires four thresholds: \emph{\$Th\_initSlot} and \emph{\$Th\_endSlot}
define the initial and the final instant of the observation time slot,
whereas \emph{\$Th\_sizeSlot} represents the width of the
sub-intervals considered in the evaluation of performance indices;
\emph{\$Th\_OpRtVar} is used to specify the upper bound for the
service response time slope between two sub-intervals.
The binding of these thresholds is intrinsically more difficult than others,
and the numerical values are set by exploiting historical data
(obtained by previous performance analysis) to accurately tune the slope
used as boundary for the increase of the response time.

Then, we apply our process to detect performance antipatterns. The pre-cal\-cu\-lus
verifies the static, dynamic, and deployment characteristics of performance antipatterns,
and pointed out that: (i) the \emph{AppServer} component may originate an instance of
the Blob antipattern in the EHS model since it (see Table \ref{tb:perfParamsBinding} and
Figure \ref{fig:caseStudy:ehs-componentDiagram}) has more than \emph{4} connections towards
other components, sends more than \emph{5} messages in the \emph{PatientInfo} service, and
is deployed on a different node with respect to the \emph{Database} component,
hence the LAN utilization need to be monitored. In this way for the Blob antipattern the
monitoring of performance indices is limited to the LAN utilization. In our case study,
$SC_0$ = \{$EHS$, \{Blob, TJ, CPS, The Ramp, More is Less\}, \{Uti\-li\-za\-tion(LAN), RT(Up\-da\-te\-Vi\-tal\-Pa\-ra\-me\-ters), \ldots \}\}.

\begin{figure}[htbp]
    \centering
        \resizebox{1\textwidth}{!}{%
        \includegraphics{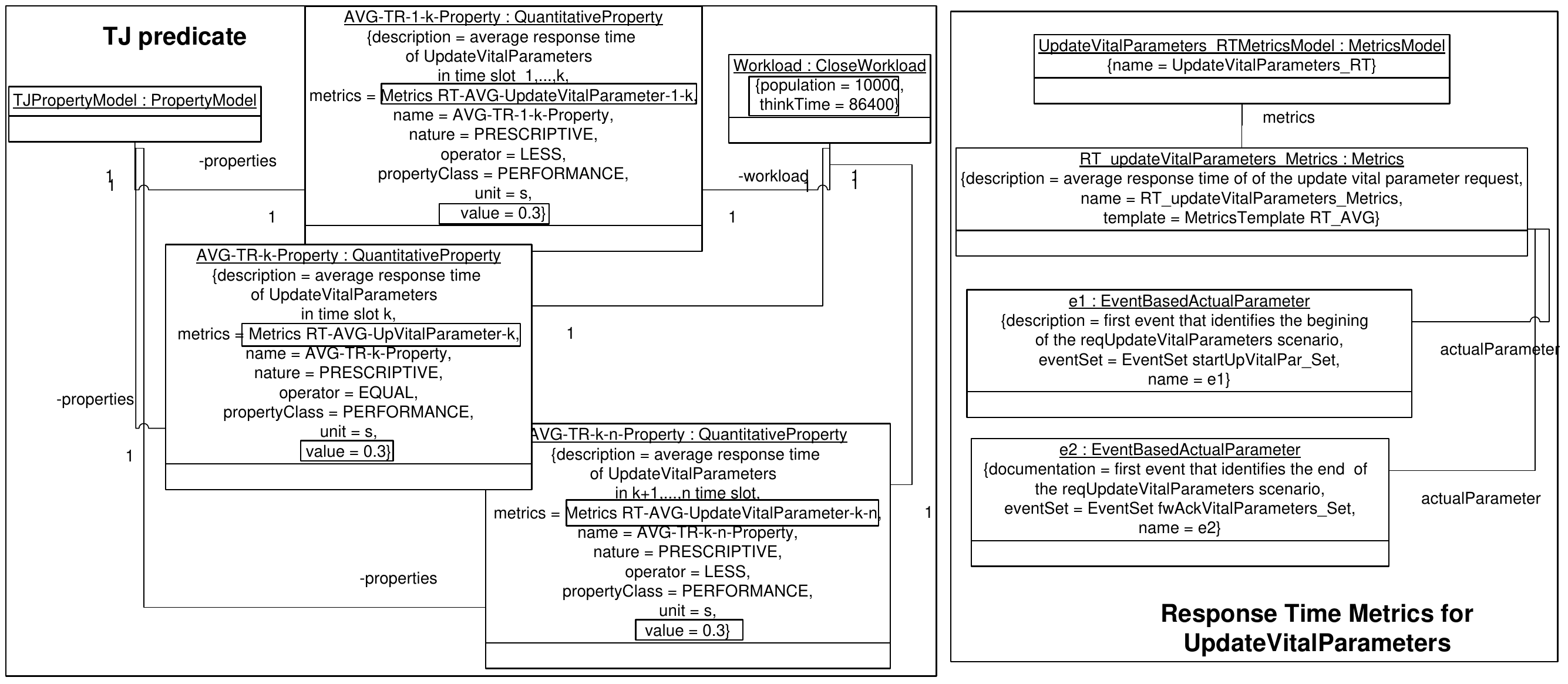}}
        \vspace{-0.6cm}
        \caption{EHS- actual property model for the TJ antipattern.}
  \label{fig:actualizedTJProperty}
\end{figure}

Our property-driven monitoring infrastructure transforms the generic property models corresponding to the pre-calculated antipatterns into the EHS actual property models. Figure \ref{fig:actualizedTJProperty} shows the actual property model for TJ performance view obtained by actualizing the generic property model used in TJ performance sub-view (see Figure ~\ref{fig:rt_template}). In particular, the actual TJ property sets the workload parameters (population to 10000 and thinking time to 86400), the threshold value to 0.3 and the average response time metrics related to the EHS \emph{UpdateVitalParameters} operation instance that is reported in the model of the right-side of the figure. The derived actual property model is sent to GLIMPSE that, invoking the translator, obtains the Drools rules, thus to instruct the probes. Then, the monitoring infrastructure starts to monitor the occurrences of the pre-calculated antipatterns.

\begin{figure}[htbp]
\vspace{-0.2cm}
    \centering
    \resizebox{1\textwidth}{!}{%
        \includegraphics{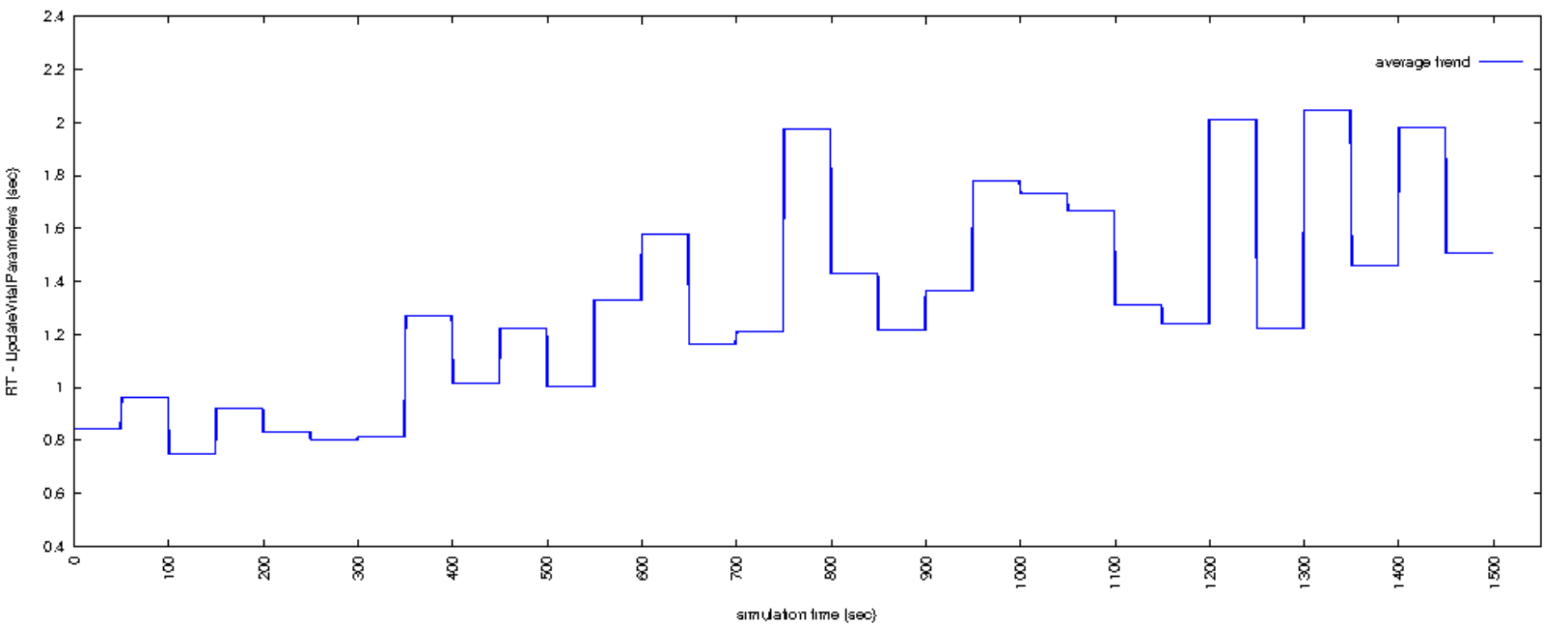}}
        \vspace{-0.7cm}
    \caption{EHS- the \emph{TJ} antipattern occurrence.}
  \label{fig:tjCaseStudy}
  \vspace{-0.5cm}
\end{figure}

In the following we report the data of RT(Up\-da\-te\-Vi\-tal\-Pa\-ra\-me\-ters) that provided the evidence for the TJ antipattern occurrence.
Figure \ref{fig:tjCaseStudy} illustrates the trend of the average response time
for the \emph{UpdateVitalParameters} service (y-axis)
over  time (x-axis) while fixing
the workload to ten thousand patients.
We obtained this trend by following the antipattern thresholds (see Table \ref{tb:perfParamsBinding}):
(i) the  time starts at the interval $0$; (ii) the size of the  intervals is fixed to
$50$ seconds; (iii) the time ends after $1500$ seconds; (iv) for each interval we calculated the
average response time of the observed completions and we verify if the slope among two intervals is larger
than 0.3 seconds. For example, in Figure \ref{fig:tjCaseStudy}
there are several subsequent intervals that violate the threshold, e.g. in the  interval
from 300 to 350 seconds the average response time is 0.81 whereas the interval from 350 to 400 seconds
the average response time is 1.27 hence the slope is 1.27 - 0.81 = 0.46 (larger than 0.3).
Hence, we can assert that the TJ antipattern occurs in the \emph{UpdateVitalParameters} service since
several subsequent intervals differ with values larger than the antipattern threshold.

The \emph{TJ} antipattern is solved by substituting the hardware platform (\emph{DbHost}) hosting the database
component with a new one whose processing power is increased by a factor of 1/100. Such refactoring action,
i.e., the replacement of the \emph{DbHost} hardware platform with a faster one, is a
\emph{hardware refactoring} (see Section \ref{sec:process}) and leads to a system configuration
$SC_{1500}$ = \{$EHS$, \{Blob, TJ, CPS, The Ramp, More is Less\}, \{Utilization(LAN), RT(Up\-da\-te\-Vi\-tal\-Pa\-ra\-me\-ters), \ldots \} \}.
where no changes have been performed to the set of antipatterns and active monitors.
The software model, refactored with the TJ antipattern, has been newly analyzed and we found
that such refactoring action was beneficial in fact the response time of \emph{UpdateVitalParameters}
service significantly improves. The requirement has been fulfilled, in fact even while considering a workload of
ten thousands patients the system offers the service in an average time of 0.5 seconds, as stated in the requirement.

\vspace{-0.3cm}
\section{Related Work}
\label{sec:related}
\vspace{-0.15cm}

Several approaches have been recently introduced to specify and detect code smells and antipatterns
\cite{DBLP:conf/icsoc/MohaPNCGBJ12,DBLP:journals/ese/KhomhPGA12}.
They range from manual approaches, based on inspection techniques \cite{Travassos-1999}, to metric-based heuristics
\cite{DBLP:conf/csmr/OlivetoKAG10}, using rules and thresholds on various
metrics \cite{DBLP:journals/tse/MohaGDM10} or Bayesian Belief Networks \cite{DBLP:journals/jss/KhomhVGS11}.

Parsons et al. \cite{DBLP:journals/jot/ParsonsM08} present a framework
for detecting performance antipatterns, where a rule-based performance diagnosis tool, named Performance
Antipattern Detection (PAD), is presented. The approach deals with Component-Based Enterprise Systems,
targeting Enterprise Java Bean (EJB) applications only. It is based on monitoring data from running systems,
it extracts the runtime system design and detects EJB antipatterns by applying rules to it. However,
the scope of \cite{DBLP:journals/jot/ParsonsM08} is restricted to such domain (i.e. EJB running applications),
and the detection of the performance problems is not allowed earlier in the development process.
In complement, our approach intends to work at the design level and it can be applied
early in the software life-cycle.

In our previous work we first tackled the problem of providing a more formal representation for performance
antipatterns \cite{SmithWilliams2003} by introducing first-order logic rules that express a set of system properties under which an
antipattern occurs \cite{antipatterns-sosym-2012}. We also started to investigate the problem of providing
QoS-based feedback by means of design and runtime knowledge in \cite{DBLP:conf/iceccs/MirandolaT12}. On the contrary,
in this paper we focus on how the runtime knowledge can be used for the detection of software performance
antipatterns.

Besides antipatterns-based approaches, there is some literature that deals with the automated generation of
architectural feedback, in particular: (i) rule-based approaches (e.g. \cite{DBLP:conf/qosa/KavimandanG09})
encapsulate general knowledge on how to improve system performance into executable rules; (ii) design space
exploration approaches (e.g. \cite{DBLP:conf/cpe/ZhengW03}) explore the design space by examining alternatives
that can cope with performance flaws; (iii) metaheuristic approaches (e.g. \cite{DBLP:conf/wosp/MartensKBR10})
make use of evolutionary algorithms looking for design alternatives aimed at improving the system performance.

\vspace{-0.55cm}
\section{Conclusions}\label{sec:conclusion}
\vspace{-0.35cm}

In this paper we presented a model-driven approach that allows
the detection of software performance antipatterns at runtime.
In particular, the process we proposed for the detection of performance
antipattern at runtime is composed by two steps: pre-calculus and monitoring.
For the second step, we proposed an adaptation of a monitoring infrastructure,
built on top of a property meta-model (PMM), that has been presented in \cite{SERENE11}.
From models conform to PMM it is possible to generate Drools Rules that
are interpreted by the generic monitoring framework GLIMPSE. In this way,
it is possible to instruct the monitoring infrastructure to concrete set up.
The benefit of our approach is the introduction of a pre-calculus step that partially evaluates the antipattern
logic formulation, checks off-line the static, dynamic, and deployment characteristics of antipatterns, and
only leaves the verification at run time of the performance view predicates to the monitoring infrastructure.

Our future work agenda is represented by the following tasks: (i) complete the library of the Generic Property Models;  (ii) extensively validate the PMM2DROOLs transformation in order to test its correctness in several scenarios; (iii) evaluate the overhead of the GLIMPSE rule engine w.r.t. compiled probes; (iv) experiment the approach on case studies coming from industrial experiences.
\vspace{-0.45cm}

\bibliographystyle{eptcs}
\bibliography{biblio}

\end{document}